\documentclass[pre,preprint]{revtex4}

\usepackage{subfig}
\usepackage{graphicx}
\usepackage{amsmath}
\usepackage{amssymb}
\usepackage{amsthm}
\usepackage{bm}
\usepackage{cancel}
\usepackage[normalem]{ulem}
\usepackage{xfrac}

\usepackage{color}

\usepackage{hyperref}
\usepackage{cleveref}
\usepackage{placeins}

\begin{document}

\graphicspath{{images/}}

\title{Enhancing Chaotic Behavior at room temperature in GaAs/(Al,Ga)As Superlattices}
\author{M. Ruiz-Garcia${}^\dagger$\footnote{Corresponding author; email: miruizg@ing.uc3m.es}, J. Essen${}^\ddagger$${}^\triangle$, M. Carretero${}^\dagger$, L. L. Bonilla${}^\dagger$, and B. Birnir${}^\triangle$${}^\circ$}
\address{${}^\dagger$Gregorio Mill\'an Institute for Fluid Dynamics, Nanoscience and Industrial Mathematics, and Department of Materials Science and Engineering and Chemical Engineering, Universidad Carlos III de Madrid,
Avenida de la Universidad 30, 28911 Legan\'es, Spain}
\address{${}^\ddagger$Department of Physics, University of California, Santa Barbara, 93106, United States}
\address{${}^\triangle${Department of Mathematics and Center for Complex and Nonlinear Science, University of California, Santa Barbara, 93106}, United States}
\address{${}^\circ$School of Engineering and Natural Science, University of Iceland, 107 Reykjav\'ik,  Iceland}

\begin{abstract}
Previous theoretical and experimental work has put forward 50-period semiconductor superlattices as fast, true random number generators at room temperature. Their randomness stems from feedback between nonlinear electronic dynamics and stochastic processes that are intrinsic to quantum transitions. This work theoretically demonstrates that shorter superlattices with higher potential barriers contain fully chaotic dynamics over several intervals of the applied bias voltage  compared to the 50-periods device which presented a much weaker chaotic behavior. The chaos arises from deterministic dynamics, hence it persists even in the absence of additional stochastic processes. Moreover, the frequency of the chaotic current oscillations is higher for shorter superlattices. These features should allow for faster and more robust generation of true random numbers.
\end{abstract}

\maketitle

\section{Introduction}

\begin{figure}[t]
    \includegraphics[height=0.3\textheight]{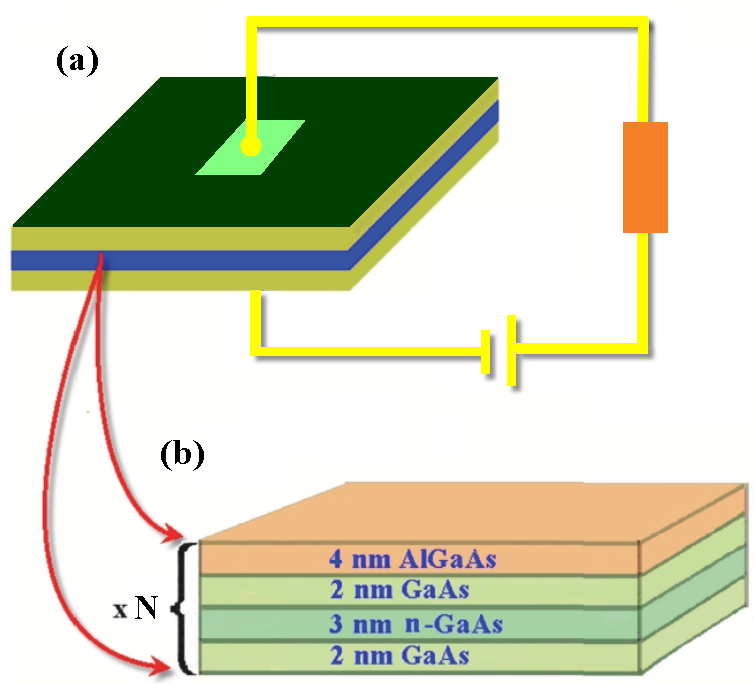}
    \caption{Simplified image of a semiconductor superlattice. A external voltage is applied between the contacts at the top and bottom of the device, which consists of $N$ periods of GaAs/Al$_x$Ga$_{1-x}$As. The 7nm GaAs wells are divided into three zones to prevent doping diffusion.}
    \label{fig:1}
\end{figure}

Fast random number generators (RNGs) are relied upon for many applications including, inter alia, data encryption systems, stochastic modeling, and secure communication \cite{sti95,gal08,nie00}. In many cases, the RNG is substituted by a numerical algorithm that produces a seemingly unpredictable sequence of numbers when a short random `seed' is entered as input \cite{asm07}. While this approach is convenient and inexpensive, the resulting number sequences are only pseudorandom, i.e. the algorithm will produce identical number sequences given identical seeds. To eliminate this vulnerability, it is necessary to find fast and reliable physical sources of entropy that produce true random number sequences. Recently, chaotic semiconductor lasers \cite{uch08,murphy08,reid09,kanter10,scia15} and superlattices \cite{li2013fast} have been used for fast generation of truly random numbers at a rate of tens or hundreds of Gb/s. In both cases, quantum fluctuations are coupled with chaotic dynamics to produce a macroscopic fluctuating signal that is detectable using conventional electronics. However, while semiconductor lasers require a mixture of optical and electronic components, semiconductor superlattices (SSLs) are entirely electronic submicron devices that are more readily integrated into complex circuits, see Figure \ref{fig:1}. Hence SSLs could be vastly useful, as the security of digital computers and networks relies on fast generation of truly random numbers.

Two different time scales are involved in the dynamics of SSLs. The inter-site tunneling and inter-subband relaxation processes occur on much shorter timescales than the dielectric relaxation processes \cite{bonilla1994,bonilla2000microscopic}. Therefore, the long timescale dynamics of semiconductor lasers \cite{scia15} and superlattices \cite{bonilla2005nonlinear,alvaro2014noise} are typically modeled using semiclassical equations, while the short timescale processes are treated stochastically.
Chaotic dynamics via period-doubling cascades have been theoretically predicted in optically-driven assymetric quantum well systems \cite{Gal96a,Bat03} and in 100-period SSLs \cite{ama02}. Until recently, experimental observation of chaos in SSLs required ultralow temperatures \cite{bonilla2005nonlinear}. Huang et al argued \cite{huang2012experimental} that phonon-assisted transport though the $X$-valley of AlAs allowed a thermal distribution of carriers to diffuse through the SSL, eliminating self-sustained oscillations and spontaneous chaos at higher temperatures.
Therefore an Aluminum concentration of 45\% was chosen for the GaAs/AlGaAs SSL in order to maximize the lowest bandgap energy (making the $X$ and $\Gamma$ band gaps equal to one another). They subsequently observed current self-oscillations and spontaneous chaos in dc-biased 50-period SSLs at room temperature for the first time \cite{huang2012experimental,huang2013spontaneous}. Weak noise-enhanced chaos has been found in simulations for 50-periods SSL \cite{alvaro2014noise}, which opened the way to new perspectives that could optimize the chaotic behavior in SSLs.

In this paper, we investigate the behavior of the sequential resonant tunneling (SRT) model for shorter SSLs at room temperature. We consider two different barrier heights corresponding to an Aluminum content of $45\%$ (as in recent experiments \cite{huang2012experimental,huang2013spontaneous}) and a different concentration of $70\%$ to study the possible effect of increasing the barrier height on the dynamical behavior. We observe a period doubling cascade to chaos on wide voltage intervals for a 10-period SSL. Moreover, the chaotic self-oscillations occur at much higher frequencies for these shorter superlattices, increasing the rate of random number generation. The outline of the paper is as follows. In section \ref{model}, we describe the SRT model of nonlinear electronic transport in SSLs. The results of our numerical simulations are reported in section \ref{sec:results}, and a discussion of our results is contained in section \ref{conclusion}.

\section{Model}\label{model}
Many phenomena are captured by means of a quasi-one-dimensional resonant sequential tunneling model of nonlinear charge transport in SSLs \cite{bonilla2005nonlinear,bonilla2002,bonilla2009nonlinear}. Consider a weakly coupled superlattice having $N$ identical periods of length $l$ and total length ${ L=N l }$ subject to a dc bias voltage $V$. The evolution  of $F_i$, the average electric field at the SSL period $i$, and the total current density, $J(t)$, is described by Ampere's law
\begin{equation}\label{eq:ST1}
J(t) = \epsilon \frac{dF_i}{dt}+J_{i\to i+1},
\end{equation}
and the voltage bias condition
\begin{equation}
    \label{eq:bias}
\sum_{i=1}^{N}F_i=\frac{V}{ l }.
\end{equation}
Fluctuations of $F_i$ away from its average value ${F_{\text{avg}}=eV/L}$ arise from the inter-site tunneling current $J_{i\to i+1}$, which appears in equation~\eqref{eq:ST1}. A microscopic derivation of $J_{i\to i+1}$ produces the result \cite{bonilla2000microscopic,bonilla2002}
\begin{equation}
J_{i\rightarrow i+1}=\frac{e n_i}{l} v^{(f)}(F_i)-J_{i\rightarrow i+1}^{-}(F_i,n_{i+1},T),
\end{equation}
in which $n_i$ is the electron sheet density at site $i$, $-e<0$ is the electron charge and $T$ is the lattice temperature. Here the forward velocity, $v^{(f)}(F_i)$, is peaked at resonant values of $F_i$ for which one or more energy levels at site $i$ are aligned with the levels at site $i+1$, and
\begin{equation}
\begin{split}
&J_{i\rightarrow i+1}^{-}(F_i,n_{i+1},T)= \\
&\ \ \frac{em^* k_B T}{\pi \hbar^2 l}v^{(f)}(F_i) \ln \left[1+ e^{-\frac{eF_i l}{k_B T}}\left( e^{\frac{\pi \hbar^2 n_{i+1}}{m^* k_B T}} -1\right)  \right]\!,
\end{split}
\end{equation}
where the reference value of the effective electron mass in Al$_x$Ga$_{1-x}$As is $m^*=(0.063 +0.083x) m_e$, and $k_B$ is the Boltzmann constant. The $n_i$ are determined self-consistently from the discrete Poisson equation,
\begin{equation}
n_i=N_D +\frac{\epsilon}{e}(F_i-F_{i-1}),
\end{equation}
where $N_D$ is the doping sheet density and $\epsilon$ is the average permittivity. The field variables $F_i$ are constrained by boundary conditions at $i=0$ and $i=N$ that represent Ohmic contacts with the electrical leads
\begin{equation}\label{eq:STN}
J_{0 \rightarrow 1}=\sigma_0 F_0, \quad J_{N \rightarrow N+1} =\sigma_0 \frac{n_{N}}{N_D} F_N,
\end{equation}
where $\sigma_0$ is the contact conductivity. Shot and thermal noise can be added as indicated in \cite{alvaro2014noise,bon2016jmi}.

\begin{table}[b]
	\centering
	\begin{ruledtabular}
	\begin{tabular}{lllllll}
		$T$ (K) & $N_D$ (cm${}^{-2}$)          & $l_b$ (nm)& $l_w$ (nm)&
		$s$ ($\mu$m) \\ \hline
		$295$& $6 \times 10^{10}$ & $4$ & $7$ &
		$60$
	\end{tabular}
    \vspace{1ex}
	\begin{tabular}{llll}
		V$_{barr}$ (meV) & E$_1$ (meV) & E$_2$ (meV) & E$_3$ (meV)\\ \hline
		$600$            & $53$     & $207$          & $440$ \\
		$388$            & $45$     & $173$          & $346$
	\end{tabular}
	\end{ruledtabular}
	\caption{(Top) The design parameters of the superlattice. (Bottom) Values of the potential barrier and energy levels for GaAs/Al$_{0.7}$Ga$_{0.3}$As and GaAs/Al$_{0.45}$Ga$_{0.55}$As superlattices, first and second row, respectively.}
	\label{tab:params}
\end{table}

Table~\ref{tab:params} gives the numerical values of the parameters used in the simulations. The GaAs/Al$_{0.45}$Ga$_{0.55}$As configuration corresponds with the configuration used in experiments \cite{li2013fast,huang2012experimental,Yin16}. The rest of the parameters are as follows: $l_b$ and $l_w$, with $l=l_b+l_w$, are the barrier and well lengths, respectively, and $A=s^2$ is the transversal area of the superlattice. The contact conductivity is a linear approximation of the behavior of $J_{0\rightarrow 1}$, which depends on the structure of the emitter; the value has been taken to reproduce the experimental results with $N=50$: $\sigma_0=0.783$ A/Vm for $V_{barr}=388$ meV ($x=0.45$) and $\sigma_0=0.06$ A/Vm for $V_{barr}=600$ meV ($x=0.7$), where $V_{barr}$ is the height of the barrier \cite{alvaro2014noise,huang2012experimental}.

\section{Results}
\label{sec:results}

\begin{figure}[ht!]
    \centering
    \textsc{Power Spectra and Bifurcation Diagram}\\
    \includegraphics[scale=1]{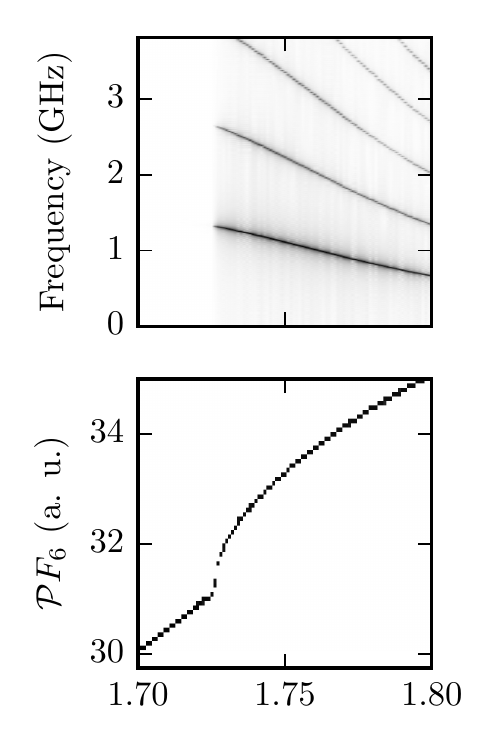}
    \hspace{-0.2in}
    \includegraphics[scale=1]{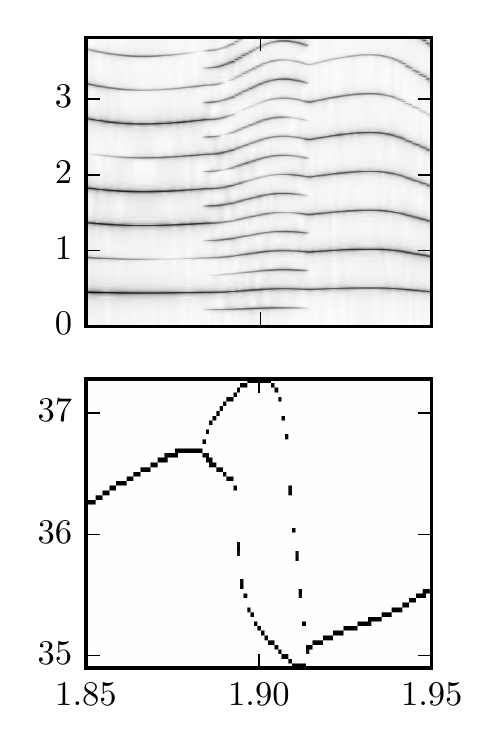}
    \hspace{-0.2in}
    \includegraphics[scale=1]{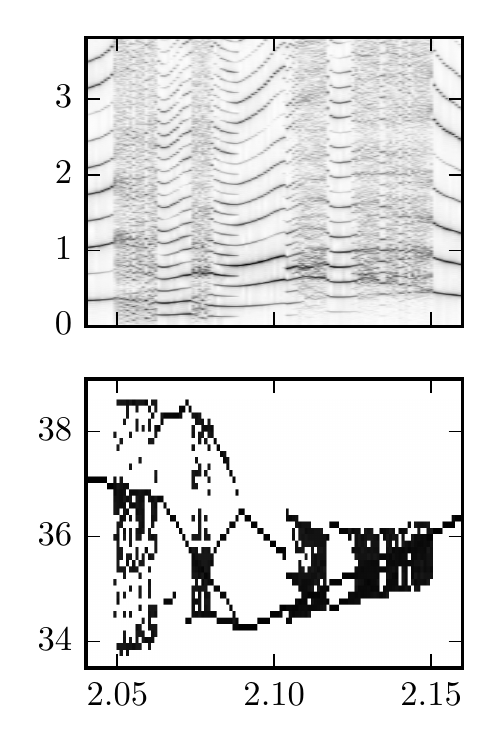}\\
    \vspace{-4ex} {\footnotesize Voltage (V)}
    \caption{ Power spectrum versus voltage and bifurcation diagrams for a 10-period GaAs/Al$_{0.7}$Ga$_{0.3}$As SSL and different voltage regions. (Top row) Power spectrum of $J(t)$ versus  voltage. (Bottom row) Bifurcation diagram of Poincar\'e map versus voltage. The Hopf bifurcation from the steady state is shown in the first column. A period doubling ``bubble'' is shown in the second column. A period-doubling cascade is shown in the third column.}
    \label{fig:bifurcation}
\end{figure}

\begin{figure}[p]
    \centering
    \includegraphics[scale=1]{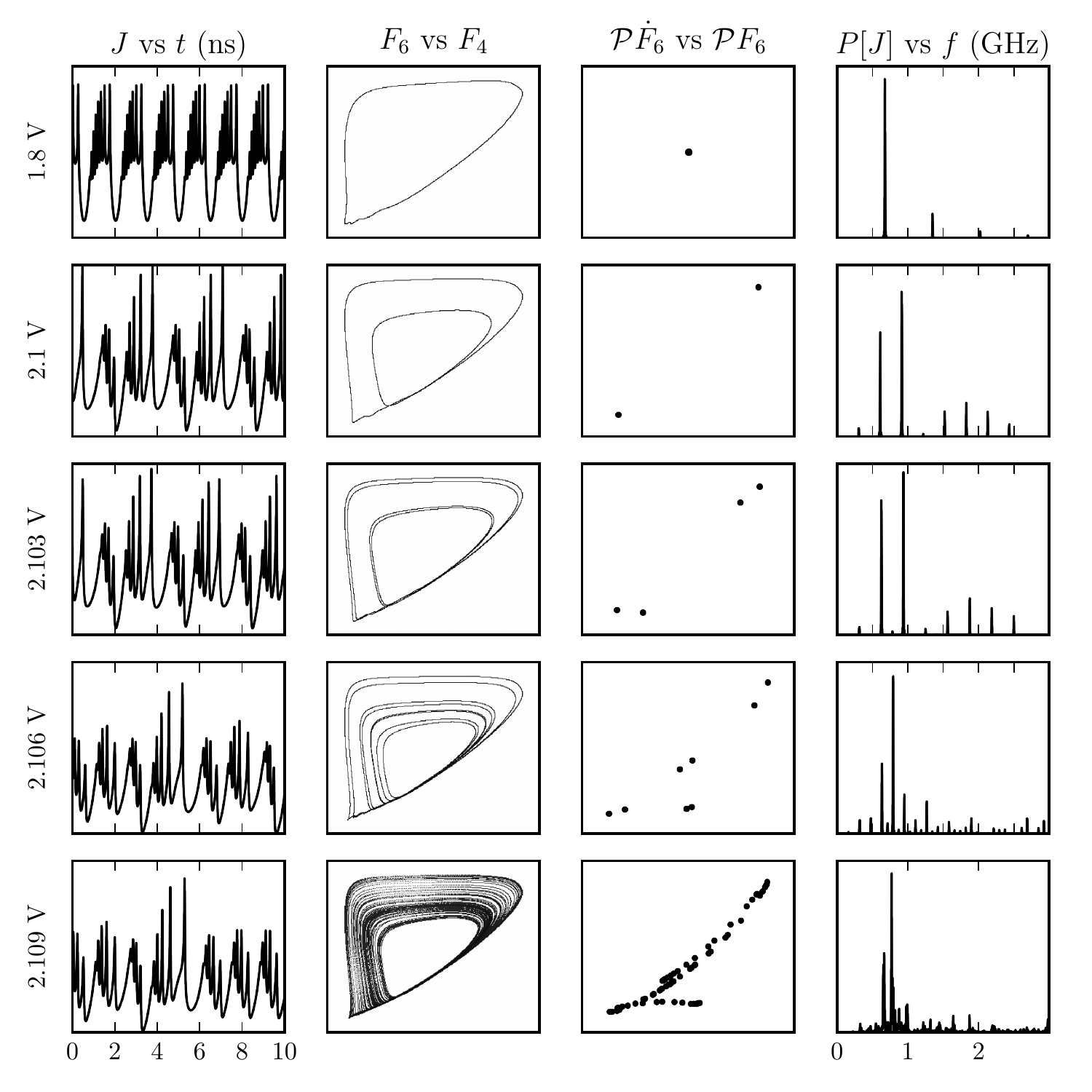}
    \caption{Representative phase portraits for the 10-period GaAs/Al$_{0.7}$Ga$_{0.3}$As SSL. The first column shows the average current $J$ plotted against time $t$. The second column shows the phase portrait $F_6(t)$ plotted against $F_4(t)$. The third column shows the Poincare map $\mathcal{P}\dot F_6(t^*)$ plotted against $\mathcal{P}F_6(t^*)$. The last column shows the power spectrum of $J(t)$. A periodic oscillation is shown in the first row. The period-doubling cascade to a chaotic attractor is shown in the bottom four rows.}
    \label{fig:portrait}
\end{figure}

We analyze the SRT model of  10-period GaAs/Al$_{0.7}$Ga$_{0.3}$As  and GaAs/Al$_{0.45}$Ga$_{0.55}$As SSLs with the material parameters indicated in table \ref{tab:params}. 
Equations~\eqref{eq:ST1}--\eqref{eq:STN} are evolved in time for $t_f=200$ ns using the forward Euler method. We remove the transient behavior due to the initial conditions by discarding first $t_i=100$ ns of evolution at each bias voltage. Bifurcations are detected via the Poincar\'e map, which is depicted in Figures~\ref{fig:bifurcation} and \ref{fig:portrait}. First, the time-evolution is projected onto a two-dimensional slice through phase space, in this case, the $F_4$-$F_6$ plane was used. When $F_4(t)$ passes through its center value, and, in order to sample the trajectory only once per cycle, $\dot F_4(t^*) < 0$, the time $t^*$ and the values of $F_6(t^*)$ and $\dot{F}_6(t^*)$ are stored. These sets of values form $\mathcal{P}F_6$ and $\mathcal{P}\dot{F}_6$.

\begin{figure}[ht!]
	\centering
	\textsc{Power Spectra and Bifurcation Diagram}\\
	\includegraphics[scale=1]{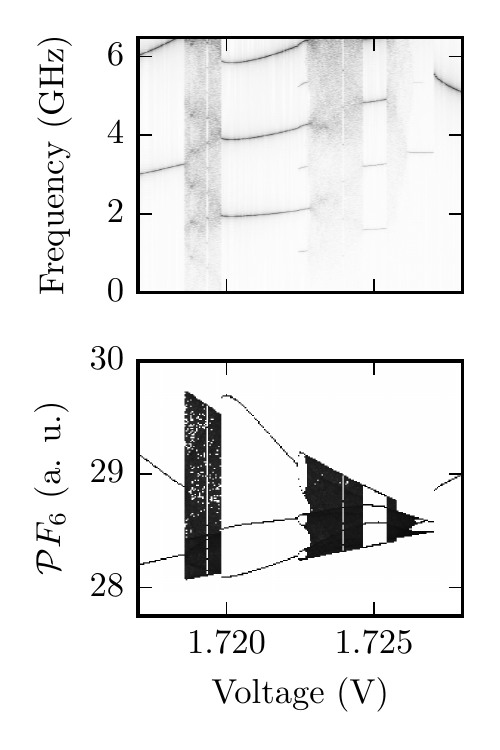}
	\caption{Power spectrum and bifurcation diagram for a 10-period GaAs/Al$_{0.45}$Ga$_{0.55}$As SSL, in a voltage region where chaotic behavior is present. (Top row) The power spectrum of $J(t)$ plotted against the bias voltage. (Bottom row) The bifurcation diagram, plotting the Poincar\'e map against the bias voltage.}
	\label{fig:bifurcation_45}
\end{figure}

The Poincar\'e map transforms the continuous time evolution in the $2N+1$-dimensional phase space (electric fields, electron densities and total current density) into a discrete map from a one-dimensional interval into itself
\cite{collet2009iterated}. Both, a stationary state and a periodic orbit will appear as a fixed point of the Poincar\'e map. A period-doubling bifurcation is identified when  one-cycles transition to two-cycles. Chaotic regions are identified where a proliferation of period-doubling bifurcations yields fractal structure in the bifurcation diagram.

We support our analysis of the Poincar\'e map by comparing our conclusions against the power spectrum
\begin{equation}
	P[J](f) = \left|\int_{t_i}^{t_f}dt\, e^{-i 2\pi ft} J(t)\right|^2,
\end{equation}
where $f$ is the frequency. As in the Poincar\'e map, different spectra are associated with different dynamical structures: (a) periodic orbits correspond to a series of peaks with widths of the same order as the frequency bin size, falling at integer multiples of the fundamental frequency, (b) period doubling bifurcations are recognized when the number of peaks in the spectrum changes by a factor of two, and a new peak appears in the power spectrum at half the fundamental frequency, (c) strange attractors have broadband spectra. These spectra may contain both sharp and broad peaks.

Figure \ref{fig:bifurcation_45} shows a voltage region where deterministic chaotic behavior is present in the simulations for the 10-period GaAs/Al$_{0.45}$Ga$_{0.55}$As SSL, see Table~\ref{tab:params}. In contrast with the $N=50$ case for the same aluminum content, there are observable windows of strong chaotic behavior, whereas  chaotic dynamics for $N=50$ appeared within very narrow voltage windows and were so weak that they became observable only by the addition of stochastic terms to the evolution equations that enhanced chaos \cite{alvaro2014noise}. Moreover, the simulations show that the lowest harmonic can reach frequencies up to $25$ GHz, at least one order of magnitude higher than those observed in the 50-period SSLs.

The bifurcation diagram for the 10-period GaAs/Al$_{0.7}$Ga$_{0.3}$As SSL, see Table~\ref{tab:params}, is presented in Figure~\ref{fig:bifurcation}, and several phase portraits are presented in Figure~\ref{fig:portrait}. Voltage windows where chaotic behavior is present are one order of magnitude wider than in the previous case, Figure \ref{fig:bifurcation_45}.
Combining the bifurcation diagram, power spectra and phase portraits of Figures \ref{fig:bifurcation} and \ref{fig:portrait}, we characterize the dynamical instabilities of the SRT model for $N=10$. At low voltages, $J(t)$ approaches a steady state.
We observe the following bifurcations:

\paragraph{Supercritical Hopf bifurcation.}
In the leftmost column of Figure~\ref{fig:bifurcation}, we observe a transition from stationary state to periodic orbit. Subsequently, we observe a circle in the phase portrait at the top row of Figure~\ref{fig:portrait}, and the power spectrum contains peaks falling at integer multiples of a fundamental oscillation frequency.

\paragraph{Period doubling bifurcation.}
In the second column of Figure~\ref{fig:bifurcation} and the second row of Figure~\ref{fig:portrait}, we observe a transition from one-cycles to two-cycles in the Poincar\`e map, so that a new peak in the power spectrum appears at half of the former fundamental frequency.

\paragraph{Period doubling cascade.}
The period doubling of the periodic orbit continues into a period-doubling cascade, resulting in a strange attractor. In particular, we have determined the first Feigenbaum constant with less than 1\% error. The rightmost column of Figure~\ref{fig:bifurcation} and the bottom three rows of Figure \ref{fig:portrait} illustrate the period-doubling cascade. Based upon the emergence of a broad peak between the two strongest harmonics, we conclude that the invariant manifold is a strange attractor.

\section{Discussion}
\label{conclusion}

This work predicts that 10-period semiconductor superlattices (SSLs), in contrast with the 50-period SSLs typically used in experiments, exhibit a more robust intrinsic deterministic chaotic behavior with faster self-sustained current oscillations. In the same direction, to increase the voltage barrier height (through increasing the aluminum content) also enhances the chaotic behavior. The deterministic chaos found in simulations of the sequential resonant tunneling (SRT) model is characterized as a Feigenbaum period doubling cascade to chaos. These results open the possibility to create faster random number generators using these shorter superlattices.

We associate the bifurcations described in Section~\ref{sec:results} with several potential applications. First, the Hopf bifurcation leads to nonlinear oscillations involving superharmonic frequencies reaching several tens of GHz. Hence these SSLs could be used as solid-state sources of electromagnetic radiation. Secondly, the half frequency found at the period doubling ``bubble,'' see the middle column of Figure~\ref{fig:bifurcation}, could be used to compress information into a desirable frequency range or to squeeze out of it undesirable noise \cite{Gr87}.

The SRT model has proven to robustly describe the essential behavior of SSLs over a wide parameter range, hence we put forward that the dynamical instabilities described in this work are the main mechanism triggering the experimentally observed chaos in SSLs. In addition, it is important to note that intrinsic quantum entropy sources are not taken into account in this work. In the real system, these quantum fluctuations are amplified by the deterministic dynamics, enabling the construction of true RNG \cite{murphy08,bon2016jmi}.

\acknowledgments
 This material is based upon work supported by, or in part by, the U. S. Army Research
Laboratory and the U. S. Army Research Office under contract/grant number
 444045-22682 and by the Ministerio de Econom\'\i a y Competitividad
of Spain under grant MTM2014-56948-C2-2-P. MRG also acknowledges support from MECD through the FPU program and from MINECO-Residencia de Estudiantes. MRG thanks UCSB Math Department for their hospitality during a stay partially supported by Universidad Carlos III de Madrid.


\begin{thebibliography}{99}
\bibitem{sti95} D.R. Stinson, \emph{Cryptography: Theory and Practice, 3rd ed.} (CRC Press, Boca Raton, 2006).

\bibitem{gal08} R.G. Gallager, \emph{Principles of Digital Communication} (Cambridge University Press, Cambridge, UK, 2008).

\bibitem{nie00} M.A. Nielsen, I.L. Chuang, \emph{Quantum Computation and Quantum Information} (Cambridge University Press, Cambridge, UK, 2000).

\bibitem{asm07} S. Asmussen, P.W. Glynn, \emph{Stochastic Simulation: Algorithms and Analysis} (Springer-Verlag, New York, 2007).

\bibitem{uch08} A. Uchida, K. Amano, M. Inoue, K. Hirano, S. Naito, H. Someya, I. Oowada, T. Kurashige, M. Shiki, S. Yoshimori, K. Yoshimura, P. Davis, Fast physical random bit generation with chaotic semiconductor lasers. Nat. Photonics {\bf 2}, 728-732 (2008).

\bibitem{murphy08} T. E. Murphy and R. Roy,  The world's fastest dice. Nat Photonics. {\bf 2}, 714-715 (2008).

\bibitem{reid09} I. Reidler,  Y. Aviad,  M. Rosenbluh, I. Kanter, Ultrahigh-speed random number generation based on a chaotic semiconductor laser. Phys Rev Lett. {\bf 103}, 024102 (2009).

\bibitem{kanter10} I. Kanter, Y. Aviad,  I. Reidler, E. Cohen, M. Rosenbluth. An optical ultrafast random bit generator. Nat Photonics. {\bf 4}, 58 (2010).

\bibitem{scia15} M. Sciamanna, K.A. Shore, Physics and applications of laser diode chaos. Nature Photonics {\bf 9}, 151-162 (2015).

\bibitem{li2013fast} W. Li, I. Reidler, Y. Aviad, Y. Y. Huang, H. Song, Y. H. Zhang, M. Rosenbluh, and I. Kanter, Fast Physical Random-Number Generation Based on Room-Temperature Chaotic Oscillations in Weakly Coupled Superlattices, Phys. Rev. Lett. \textbf{111}, 044102 (2013).

\bibitem{bonilla1994} L. L. Bonilla, J. Gal\'{a}n, J. A. Cuesta, F. C. Mart\'{\i}nez and J. M. Molera, Dynamics of electric field domains and oscillations of the photocurrent in a simple superlattice model. Phys. Rev. B {\bf 50}, 8644 (1994).

\bibitem{bonilla2000microscopic} L. L. Bonilla, G. Platero, and D. S\'anchez, Microscopic derivation of transport coefficients and boundary conditions in discrete drift-diffusion models of weakly coupled superlattices.  Phys. Rev. B  {\bf 62}, 2786 (2000).

\bibitem{bonilla2005nonlinear} L. L. Bonilla and H. T. Grahn, Non-linear dynamics of semiconductor superlattices. Reports on Progress in Physics {\bf 68}, 577 (2005).

\bibitem{alvaro2014noise} M. Alvaro, M. Carretero, and L. Bonilla, Noise-enhanced spontaneous chaos in semiconductor superlattices at room temperature.  EPL (Europhysics Letters)  {\bf 107}, 37002 (2014).

\bibitem{Gal96a} B. Galdrikian and B. Birnir, Period Doubling and Strange Attractors in Quantum Wells. Phys. Rev. Lett. {\bf 76}, 3308 (1996).

\bibitem{Bat03} A. A. Batista, B. Birnir, P. I. Tamborenea and D. S. Citrin, Period-doubling and Hopf bifurcations in far-infrared driven quantum well intersubband transitions. Phys. Rev. B {\bf 68}, 035307 (2003).

\bibitem{ama02} A. Amann, J. Schlesner, A. Wacker, E. Sch\"oll, Chaotic front dynamics in semiconductor superlattices. Phys. Rev. B {\bf 65}, 193313 (2002).

\bibitem{huang2012experimental} Y. Huang, W. Li, W. Ma, H. Qin, and Y. Zhang, Experimental observation of spontaneous chaotic current oscillations in GaAs/Al0.45Ga0.55As superlattices at room temperature.  Chinese Science Bulletin  {\bf 57}, 2070 (2012).

\bibitem{huang2013spontaneous} Y. Huang, W. Li, W. Ma, H. Qin, H. T. Grahn, and Y. Zhang, Spontaneous quasi-periodic current self-oscillations in a weakly coupled GaAs/(Al,Ga)As superlattice at room temperature.  Applied Physics Letters  {\bf 102}, 242107 (2013).

\bibitem{bonilla2002} L. L. Bonilla, Theory of Nonlinear Charge Transport, Wave Propagation and Self-oscillations in Semiconductor Superlattices. Journal of Physics: Condensed Matter {\bf 14}, R341 (2002).

\bibitem{bonilla2009nonlinear} L. L. Bonilla and S. W. Teitsworth, \emph{Nonlinear wave methods for charge transport}. (Wiley VCH, Weinheim, 2009).

\bibitem{bon2016jmi} L.L. Bonilla, M. Alvaro, and M. Carretero, Chaos-based true random number generators. Journal of Mathematics in Industry {\bf 7}, 1 (2016).

\bibitem{Yin16} Z. Yin, H. Song, Y. Zhang, M. Ruiz-Garcia, M. Carretero, L. L. Bonilla, K. Biermann and H. T. Grahn, Noise-enhanced chaos in a weakly coupled GaAs/(Al,Ga)As superlattice, Phys. Rev. E {\bf 95}, 012218 (2017).

\bibitem{collet2009iterated} P. Collet and J. Eckmann, \emph{Iterated Maps on the Interval as Dynamical Systems}, Modern Birkh{\"a}user Classics (Birkh{\"a}user Boston, 2009).

\bibitem{Gr87} R. Graham, Squeezing and frequency changes in harmonic oscillations. Journal of Modern Optics  {\bf 34}, 873 (1987).


\end{thebibliography}
\end{document}